\documentclass{article}

\usepackage{arxiv}

\usepackage[utf8]{inputenc} 
\usepackage[T1]{fontenc}    
\usepackage{hyperref}       
\usepackage{url}            
\usepackage{booktabs}       
\usepackage{amsmath,amsfonts}       
\usepackage{nicefrac}       
\usepackage{microtype}      
\usepackage{lipsum}
\usepackage{graphicx}
\graphicspath{ {./images/} }
\usepackage{algorithmic}
\usepackage{xcolor}
\usepackage{stfloats}
\usepackage{url}
\usepackage{adjustbox}

\title{Impact of opinion dynamics on recurrent pandemic waves: balancing risk aversion and peer pressure}

\author{
 Sheryl L. Chang  \textsuperscript{*}\\
  Centre for Complex Systems\\
  Faculty of Engineering\\
  The University of Sydney, Australia \\
  \texttt{sheryl.chang@sydney.edu.au} \\
   \And
 Quang Dang Nguyen\textsuperscript{*} \\
  Centre for Complex Systems\\
  Faculty of Engineering\\
  The University of Sydney, Australia\\
  \And
 Carl J. E. Suster \\
  Centre for Infectious Diseases and 
  \\Microbiology–Public Health\\
  Westmead Hospital\\
  NSW, Australia\\
  \And
  Christina M. Jamerlan \\
  Centre for Complex Systems\\
  Faculty of Engineering\\
  The University of Sydney, Australia\\
  \And 
  Rebecca J. Rockett \\ 
  Centre for Infectious Diseases and \\
  Microbiology–Public Health\\
  Westmead Hospital\\
  NSW, Australia\\
  \And 
  Vitali Sintchenko\\
  Centre for Infectious Diseases and \\
  Microbiology–Public Health\\
  Westmead Hospital\\
  NSW, Australia\\
  \And 
  Tania C. Sorrell \\
  Centre for Infectious Diseases 
  \\and Microbiology–Public Health\\
  Westmead Hospital\\
  NSW, Australia\\
  \And 
  Alexandra Martiniuk \\
  Faculty of Medicine and Health\\
  The University of Sydney, Australia\\
  \And
  Mikhail Prokopenko\\
  Centre for Complex Systems\\
  Faculty of Engineering\\
  The University of Sydney, Australia\\
}

\begin{document}
\maketitle
\begin{abstract}
Recurrent waves which are often observed during long pandemics typically form as a result of several interrelated dynamics including public health interventions, population mobility and behaviour, varying disease transmissibility due to pathogen mutations, and changes in host immunity due to recency of vaccination or previous infections. Complex nonlinear dependencies among these dynamics, including feedback between disease incidence and the opinion-driven adoption of social distancing behaviour, remain poorly understood, particularly in scenarios involving heterogeneous population, partial and waning immunity, and rapidly changing public opinions. This study addressed this challenge by proposing an opinion dynamics model that accounts for changes in social distancing behaviour (i.e., whether to adopt social distancing) by modelling both individual risk perception and peer pressure. The opinion dynamics model was integrated and validated within a large-scale agent-based COVID-19 pandemic simulation that modelled the spread of the Omicron variant of SARS-CoV-2 between December 2021 and June 2022 in Australia. Our study revealed that the fluctuating adoption of social distancing, shaped by individual risk aversion and social peer pressure from both household and workplace environments, may explain the observed pattern of recurrent waves of infections. 
\end{abstract}

\textsuperscript{*}: \textit{These authors contributed equally}\\

\section{Introduction}
The presence of recurrent waves of infections~\cite{Morens2009,zhang2021second} has been reported in pandemics of several pathogens~\cite{recurrent_waves,dengue,measles}. For influenza, multiple waves have been noted as a distinguishing feature of pandemics, and an opportunity for informing public health responses~\cite{Miller2009}. Accordingly, there is much interest in understanding contributing factors that can explain the observed pattern of waves~\cite{model2,Mummert2013,Xu2020,model1}. Seasonal effects, evolutionary dynamics of the pathogen, and interactions with other co-circulating pathogens are each important, however many of these explanatory models focus on the impact of human behaviour. The contact patterns of infected hosts with immunologically naive populations often determine early pandemic dynamics. As a pandemic progresses and populations observe the impacts of the disease and of control measures, it can be expected that host behaviours as a result of opinions will play an increasing role.

A pattern of recurrent waves of infections was observed during the COVID-19 pandemic, in Australia between December 2021 to June 2022 the Omicron variant was dominant. During this period, the population had a relatively high vaccine coverage (i.e., over 90\%  within adult population)~\cite{vac_coverage,porter2022new,macintyre2022modelling}. This Omicron wave of infections resulted in a a prominent peak reaching ~105,000 new cases per day around early January 2022 (over 4,100 cases per million), followed by two subsequent smaller peaks reaching ~66,000 new cases per day around early April 2022 (over 2,500 cases per million) and ~59,000 new cases per day around late May 2022 (approximately 2,300 cases per million)~\cite{covid19data}. The persistence of the Omicron variant has been partially attributed to fluctuating adoption of non-pharmaceutical interventions (NPIs), particularly social distancing (SD), during this period.

Using our census-calibrated agent-based model (ABM), we previously identified a plausible sequence of step changes in the fraction of the population adopting social distancing (i.e., SD profile) which reproduced the observed pattern of recurrent waves over a period of approximately 29 weeks when Omicron variant was dominant~\cite{chang_persistence_2023}. The SD profile was determined retrospectively by tuning the step changes to minimise the difference between simulated and actual disease incidence. The resultant profile included an initial period of low SD adoption fraction ($30\%$ of the population), followed by a period of relatively high SD adoption ($70\%$ of the population) when the incidence approached the first peak, a rapid decline to a moderate level of SD adoption (by $50-60\%$ of the population) during the second wave, and a drop to low SD-adoption (at $30\%$) around the third mid-year wave. We hypothesised that declining SD adoption could be attributed to pandemic fatigue~\cite{who_fatigue}. Although this study revealed an important relationship between social distancing adoption and pandemic spread, the underlying opinion formation mechanism and contributing factors shaping social behaviour were not explored.

In this study, we aimed to develop a model of opinion formation and dynamics which can shape complex SD adoption behaviours without relying on post hoc fitting to observation. Previous studies have modelled opinion dynamics using two perspectives: (i) opinion formation based on personal beliefs (e.g., individual risk evaluation due to fear of infection~\cite{Ferguson2007,DelValle2013,Funk2007}, or perception fatigue~\cite{Meacci2021}); and (ii) social influence (e.g., imitation~\cite{Garofalo2018,Ye2019}, averaging opinions \cite{opinion_2022}, or majority-following opinions~\cite{JAVARONE201419}). Our opinion formation model incorporates both of these aspects into our pandemic ABM. At each simulated cycle, the current opinion of each agent depends on their perceived level of risk (risk aversion) and on the opinions of the agents they interact with (peer pressure). An agent's opinion determines its behaviours on SD adoption. We quantified the impact balancing risk aversion and peer pressure on opinion dynamics, and explored how a combination of the individual and social perspectives can reproduce the recurrent pandemic waves. In doing so, we identified the effect on opinion formation of varying memory horizon, perception fatigue, and social contexts.

The remainder of the paper is organised as follows. We describe the ABM, including the pandemic transmission and control, i.e., NPIs and vaccination rollout (Section \ref{method:model}), and the opinion dynamics model  (Section \ref{method:opinion}). We then report the key factors contributing to opinion dynamics that generate recurrent pandemic waves (Section \ref{sec:results}). Finally, we discuss the limitations and implications of our study (Section \ref{sec:discussion}).

\section{Methods}
\subsection{Census-calibrated Agent-based Model}
\label{method:model}
We simulate the viral transmission of the Omicron variant of COVID-19 in Australia, as well as pandemic control measures (NPIs and vaccination roll-out), using a high-resolution agent-based model which was previously calibrated and validated~\cite{chang_modelling_2020, chang_simulating_2022,chang_persistence_2023,zachreson_how_2021,BMC}. Our model comprises 25.4 million anonymous agents, each assigned several demographic attributes derived from the latest 2021 Australian Census \cite{ABS_census}, so that the artificial population represents the key demographic characteristics of Australia. These demographic attributes, including age, gender, residential area, student enrollment, and workforce/educational groups, also determine the social  contexts in which agent-to-agent interactions and thus, disease transmission, take place~\cite{BMC,nguyen_general_2022}.

In our model, disease transmission follows a discrete-time simulation, updating each agent's health state over time: Susceptible, Latent, Infectious (asymptomatic or symptomatic), and 
Recovered. The simulation begins with infections seeded around international airports (mostly in metropolitan areas) and continues to spread within the artificial population through interactions within multiple social contexts. These interactions are simulated in two phases per workday with different contact and transmission rates: ``daytime'' cycle where interactions take place in workplace or education contexts (e.g., class, grade, school), and ``nighttime'' cycle where interactions take place in residential context (e.g., household, household cluster, neighbourhood and community).
On weekends, interactions are assumed to occur solely within residential contexts, consisting of two ``nighttime'' cycles instead~\cite{BMC}.

The probability of an exposed agent becoming infected is adjusted by checking if the agent is (i) vaccinated prior to the start of the pandemic wave, and/or (ii) adopting or complying with non-pharmaceutical interventions (NPIs), such as social distancing, case isolation, and so on. Out of all infections, only a fraction is assumed to be detected, to match the voluntary self-reporting system adopted in Australia during the Omicron stage. 
Here, we approximate the case detection rate based on the prevalence of anti-nucleocapsid antibodies in Australia \cite{lifeblood2022seroprevalence} (see Appendix for more details).

 Following previous studies \cite{zachreson_how_2021,chang_simulating_2022,chang_persistence_2023, BMC}, we assumed a high level of preemptive vaccination coverage in the population ($90\%$) with two types of vaccines (priority vaccine, with a higher efficacy; and general vaccine, with a lower efficacy) distributed prior to the start of the simulation (i.e., the start of the Omicron stage in December 2021), in line with the reported vaccination coverage and vaccine distribution in Australia \cite{vac_coverage}. In addition, we introduced a waning immunity (from both vaccination and previous infections) into the model by tracking the vaccination and infection record of each agent, with the immunity declining over time (see Appendix). Consequently, in this extension of our model, agents can be re-infected after recovering.
 
NPI adoption reduces disease transmission by reducing the strength of interactions between agents across various social contexts. These NPIs include: case isolation (CI, affecting symptomatic infectious agents and detected asymptomatic agents), home quarantine (HQ, affecting household members of infected agents), social distancing (SD, affecting susceptible agents), and school closures (SC, affecting school-aged agents, their households and teachers). Given our focus on the interplay between opinion dynamics and recurrent pandemic waves, we model SD compliance as a voluntary, opinion-driven decision which may change as the pandemic progresses (while assuming static compliance with other NPIs). Here, SD is modelled as a generic intervention which reduces the overall intensity of agent interactions, intended to capture the combined effect of reduced travel, mask-wearing, and physical distancing (see Section \ref{method:opinion}).

\subsection{Opinion dynamics}
\label{method:opinion}
\begin{figure}[!t]
    \centering
    \includegraphics[width=0.8\columnwidth]{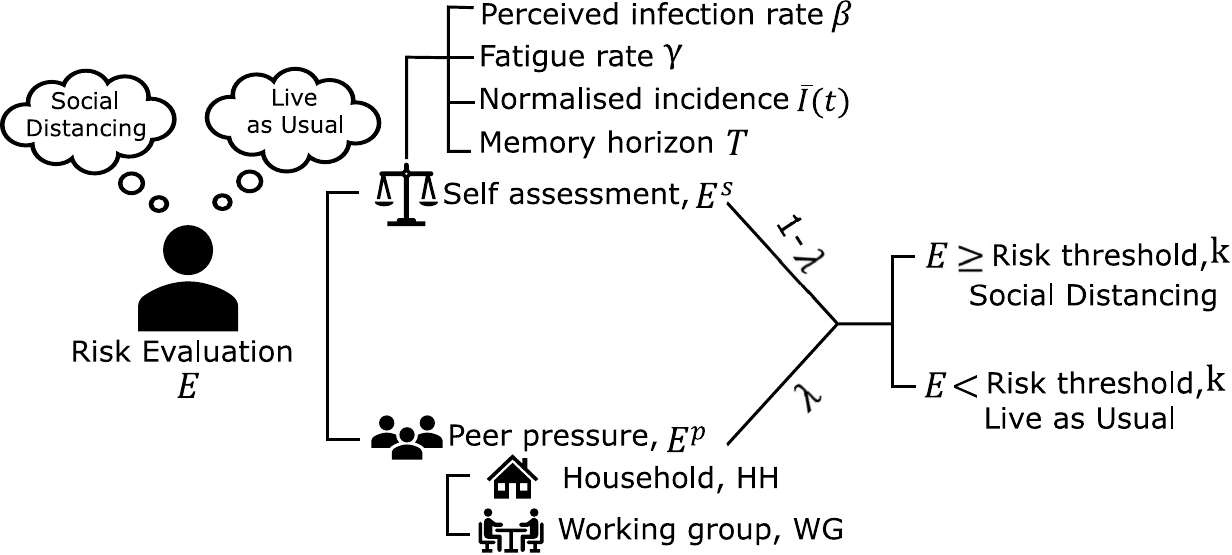}
    \caption{Opinion dynamics model overview.}
    \label{fig:method}
\end{figure}

We divide the population into three groups of agents: (i) committed, compliant agents which always comply with SD (25\% of the population, randomly assigned); (ii) non-compliant agents which never comply with SD, due to being either a contrarian or an essential worker (25\% of the population, randomly assigned); and (iii) rational agents which change their SD adoption based on the current pandemic severity (the remaining 50\% of the population). A rational susceptible agent has a choice between two behaviours: (i) ``live as usual" behaviour, and (ii)  ``socially distancing" behaviour. A rational agent may choose to switch between these two behaviours at any time as a result of the risk evaluation process following three steps: (i) self-evaluation where the opinion on SD adoption is formed individually by assessing the risk of infection (see Section \ref{method:self});  (ii) partial peer pressure where each agent may be partially influenced by opinions of the agents in their social contexts (see Section \ref{method:peer}); and (iii) integration of the self-evaluation and peer pressure steps using a weighted average (see Fig.~\ref{fig:method}).
\subsubsection{Self-evaluation}
\label{method:self}
In self-evaluation, the susceptible rational agents form their opinions on social distancing following a simple risk evaluation by comparing the perceived risk of infection, $E^\textrm{s}_i$, with a risk threshold $k$ \footnote{This study uses the conventions that superscripts on symbols are labels and subscripts are agent-specific indices.}. When the perceived risk of infection exceeds the risk threshold ($E^\textrm{s}_i \geq k$), agents adopt social distancing by reducing their interactions within the relevant social contexts (see Appendix). Conversely, when the perceived risk of infection is lower than the threshold ($E^\textrm{s}_i < k$), agents choose to live as usual and do not adjust their interactions. The perceived risk of infection for a rational agent $i$, $E^\textrm{s}_i$, follows a dynamic process dependent on the incidence record on day $d$: 

\begin{equation}
    E^\textrm{s}_i(d, T)=1-(1-\beta)^{\overline{I}(d,T)}
    \label{eq:risk}
\end{equation}
where $\beta \in [0,1]$ is the perceived probability of infection, and $\overline{I}(d,T)$ is the normalized moving average of recent daily incidence data:
\begin{equation}
        \overline{I}(d,T) =\frac{N^{\textrm{c}}}{N^{\textrm{n}}}  \frac{1}{T} \sum_{\tau=0}^{T-1} I(d-1-\tau)  \label{eq:normalised_inc}   
\end{equation}
$\overline{I}(d,T)$ averages the daily incidence, $I(d)$, between day $(d-T)$ and day $(d-1)$. $T$ acts as a memory horizon, limiting the effect of earlier incidence data on an agent's behaviour.
$N^\textrm{c}/N^\textrm{n}$ is the relative size of a typical community, defined by the ratio between the typical community size $N^\textrm{c}=10^3$ and the national population size $N^\textrm{n}=25.4 \times 10^6$, according to 2021 census \cite{ABS_census}.  Informally, this normalisation converts the global incidence to the local level at which agents perceive the disease severity.
Parameter $\beta$ quantifies the agent's attitude towards risk aversion. A higher $\beta$ leads to higher $E^\textrm{s}$, so that agents become more likely to adopt social distancing.

\begin{figure}[!t]
\vspace{1em}
    \centering
\includegraphics[width=0.55\linewidth,trim={3cm 10cm 3cm 12cm}]{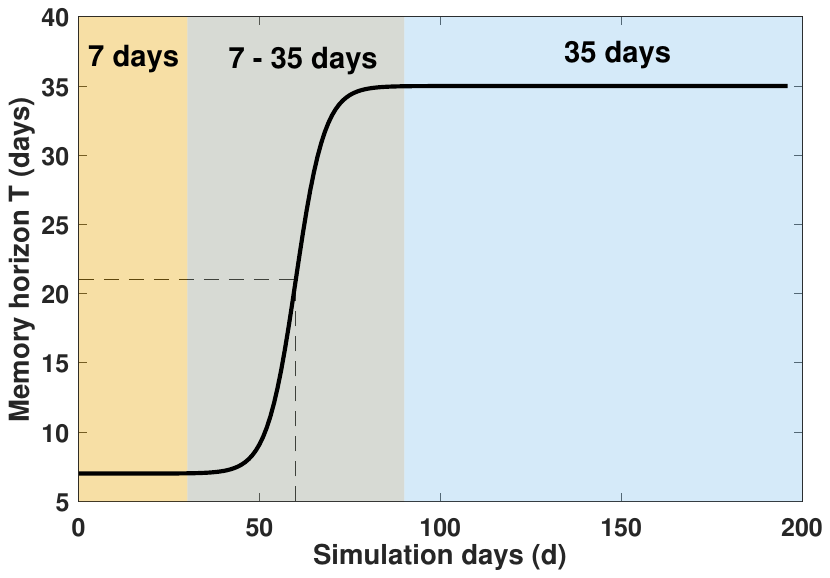}
    \caption{Sigmoid function for varying memory horizon, $T(d)$, in Eq. \ref{eq:sigmoid}. Shaded areas show the initial period with memory horizon of 7 days (yellow colour), the intermediate period of increasing value (grey), and the period with the final maximum value of 35 days (blue). Dashed lines mark the midpoint value of 21 days reached on simulation day $d = 60$.}
    \label{fig:sigmoid}
\end{figure}

In the baseline model, $T$ is a fixed parameter. We also consider an extended model where $T$ is replaced by $T(d)$, which varies globally throughout the pandemic simulation to account for the difference between time scales in transmission (i.e., faster scale) and behavioural change (i.e., slower scale). At the start of the pandemic, $T(d)$ is shorter due to the lack of information on the past incidence, so that the agents use their immediate experience to evaluate the risk of infection. As the disease spreads, agents use a longer memory horizon for risk evaluation. The evolution of the memory horizon is modelled using a sigmoid function:  

\begin{equation}
    T(d)=\frac{u}{1+e^{-v(d-L)}}+T(0)
    \label{eq:sigmoid}
\end{equation}
where $T_0$ is the baseline memory horizon length at the beginning of the simulation, $L$, determines the simulation day when the memory horizon reaches its midpoint value, and $u$ and $v$ determine the shape of the transition region in the sigmoid function. We fixed these parameters to $T(0)=7$, $L=60$, $u=28$, and $v=0.25$, resulting in $T(d)$ ranging from 7 to 35 days as illustrated in Fig. \ref{fig:sigmoid}.

Finally, we consider the effect of pandemic fatigue, with the public attention to pandemic incidence declining over time \cite{trends-sep-2022,who_fatigue}, so that the individuals gradually become less alert to the current pandemic spread and consequently, less motivated to adopt social distancing. In the baseline model the parameter $\beta$ is fixed. We consider an extended model in which we reduce the risk aversion over time by replacing $\beta$ with $\beta(d)$ in Eq. \ref{eq:risk}, governed by the perception fatigue rate~$\gamma$: 
\begin{equation}
    \beta(d) = \beta_0(1-\gamma)^d
    \label{eq:fatigue}
\end{equation}
where $\beta_0$ is the initial value of the perceived probability of infection at the start of the simulation. The inset in Fig. \ref{app:fig_mobility} (b) in Appendix shows how $\beta(d)$ declines over time with $\gamma=0.0002$. Note that for $\gamma = 0$ the time dependency is removed since $\beta(d) = \beta_0$, recovering the initial constant form. Henceforth we assume that $\beta = \beta(d)$ and distinguish the baseline model by setting $\gamma = 0$.

\subsubsection{Partial peer pressure}
\label{method:peer}

Individual opinion formation is known to be affected by opinions of other members of the community. Individuals tend to filter and integrate the information received, and align their opinions with those of their contacts in close social circles \cite{social_infleunce, social_circle}. The peer pressure is modelled by accounting for the average opinion of an agent's social context. We compute the perceived risk of infection of an agent $i$ under peer pressure ($E^\textrm{p}_i$) as the weighted sum of the average opinions in relevant opinion-affecting social context:
\begin{equation}
    E^\textrm{p}_i(d, T)=\sum_{g \in G_i}{\psi_g
 \frac{1}{|A_g|-1} \sum_{j \in A_g\backslash\{i\}}} {E^\textrm{s}_j(d, T)}
    \label{eq:social}
\end{equation}
where ${E^\textrm{s}_j}(d)$ as before is the perceived risk of infection of agent $j$ who belongs to the same context $g$ as agent $i$, $A_g\backslash\{i\}$ is the set of all agents in context $g$ excluding agent $i$, and $\psi_g \in [0,1]$ is the relative weight of the context $g$ in all social contexts $G_i$ that agent $i$ belongs to, satisfying $\sum_{g \in G_i}\psi_g =1$. In our study, we considered two representative types of opinion-affecting social contexts: households (HH) and working groups (WG). 

Following \cite{opinion_2022}, we model a combined perceived risk of infection quantified as a weighted sum of both self-evaluation ($E^\textrm{s}_i$, Eq. \ref{eq:risk}) and peer pressure ($E^\textrm{p}_i$, Eq. \ref{eq:social}) components, in accordance with the Friedkin-Johnsen model \cite{Friedkin1990}:
\begin{equation}
    E_i(d) = \lambda E^\textrm{p}_i(d) + (1-\lambda) E^\textrm{s}_i(d)
    \label{eq:combined}
\end{equation}

where $\lambda$ $\in [0,1]$ is the weight of peer pressure and $1-\lambda$ is the weight of self-evaluation in forming the current opinion.

\section{Results}
\label{sec:results}
We present results in two parts, following the risk evaluation steps: (i) self-evaluation only (Section \ref{result:SE}), and (ii) both self-evaluation and partial peer pressure (Section \ref{result:PP}).
\subsection{Self-evaluation}
\label{result:SE}
While simulating the opinion dynamics formed only by using self-evaluation, we examined three scenarios: SE1, the baseline model; SE2, where only the memory horizon varies with time, i.e. $T = T(d)$ and $\gamma = 0$; and SE3, where both the memory horizon and the perceived probability of infection vary globally with time, i.e. $T = T(d)$ and $\gamma = 0.002$. 

\begin{figure}[!t]
    \centering
    \vspace{3em}
    \includegraphics[width=0.7\linewidth,trim={4cm 8cm 4cm 9cm}]{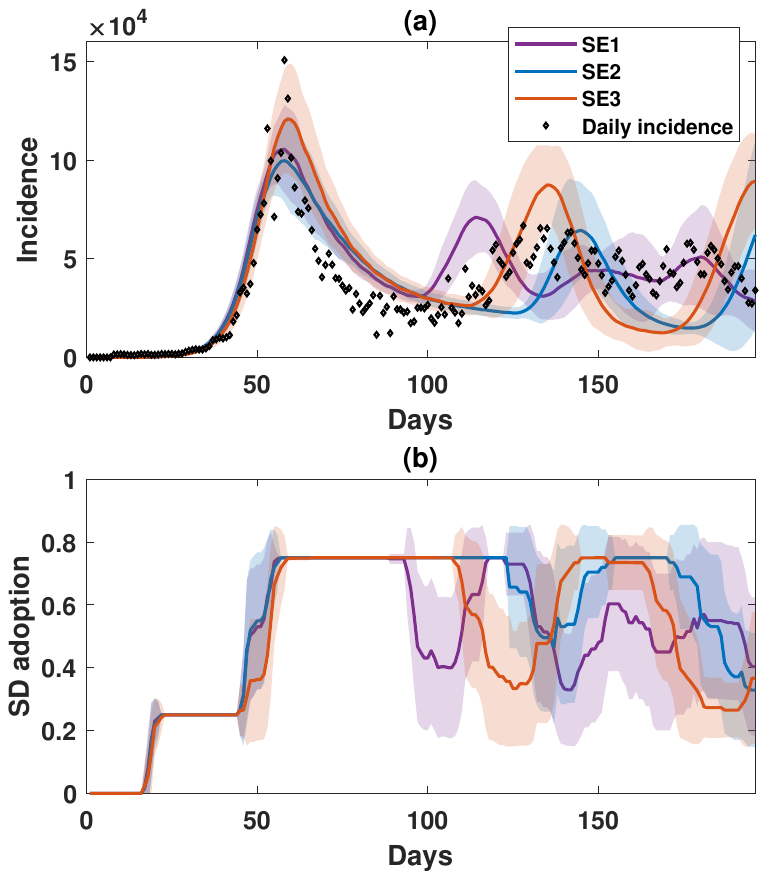}
    \vspace{1em}
    \caption{A comparison of simulated incidence (a) and SD-adoption fraction (b), with the opinion dynamics produced by self-evaluation only (SE1, baseline in purple; SE2, with changing memory horizon in blue; and SE3, with changing horizon and fatigue in orange). Refer to Table \ref{tab:sim_para} for parameterisation. Shaded areas around the solid line show standard deviation. Each simulated profile is averaged over 25 runs. The actual daily incidence between December 2021 to June 2022 is shown in scattered diamonds. Note that SE1 is equivalent to the purple profile in Fig. \ref{app:fig_beta}.}
    \label{fig:SE}
\end{figure}

For SE1, we tested different values of the constant $\beta$ to explore its effects on opinion dynamics and pandemic spread (Fig.~\ref{fig:SE} and Appendix Fig.~\ref{app:fig_beta}). We found that a moderate value of $\beta=0.5$ produced some recurrent waves, however varying $\beta$ alone could not generate a satisfactory match the observed magnitude of the first peak and the interval between waves.

Introducing the time-dependent memory horizon (SE2: Fig. \ref{fig:SE}), we observed improved accuracy in reproducing the second wave, but the difference with the first incidence peak could not be reconciled. We found that the combined effect of the time-dependent memory horizon and pandemic fatigue (SE3: Fig. \ref{fig:SE}) could reproduce the first peak and the timing of the second wave, but not the amplitude of the second incidence peak. These results suggested that the complex pandemic dynamics (e.g., recurrent waves with a prominent first wave and slower subsequent waves) observed during the Omicron stage in Australia could not be reproduced by our opinion dynamics model based solely on self-evaluation of risks.
\subsection{Partial peer pressure}
\label{result:PP}
To examine the peer pressure component coupled with scenario SE3, we varied the peer pressure weight $\lambda$, while selecting different social contexts. Specifically, we modelled three scenarios where peer pressure originated from household (PP1, $\psi^\textrm{HH}=1$), working group (PP2, $\psi^\textrm{WG}=1$), and both household and working group with equal influence (PP3, $\psi^\textrm{HH}=\psi^\textrm{WG}=0.5$). Parameters setting in these scenarios are summarised in Table \ref{tab:sim_para}. 

\begin{figure}[!t]
\vspace{4em}
    \centering
    \includegraphics[width=0.7\linewidth,trim={4cm 8cm 4cm 9cm}]{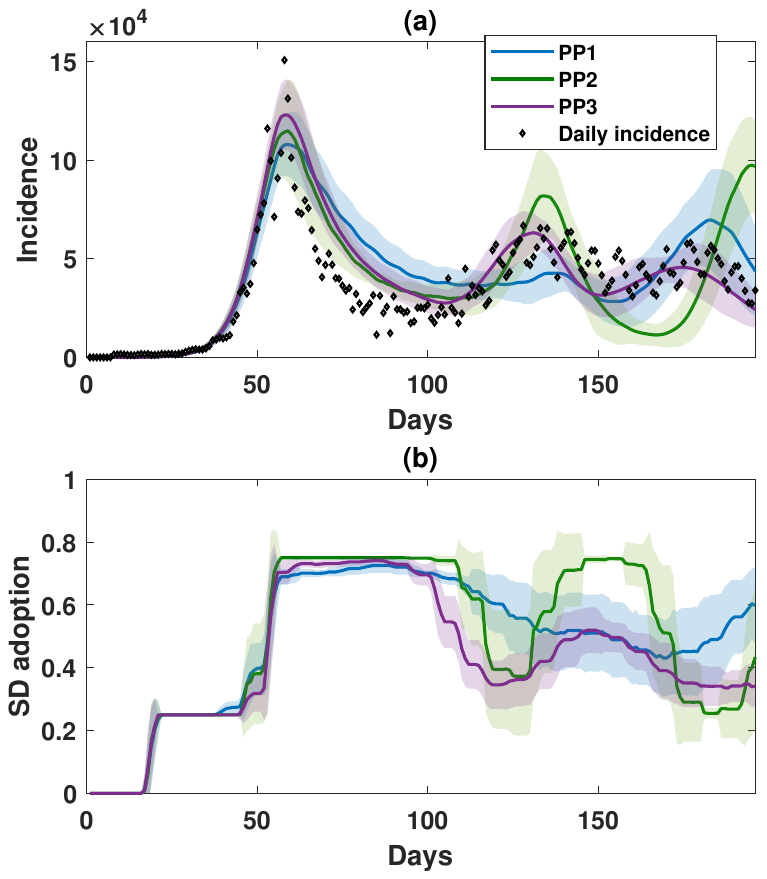}
    \vspace{1em}
    \caption{A comparison of (a) simulated incidence and (b) SD-adoption fraction, with opinions produced by combining self-evaluation and peer pressure in different social contexts (PP1, household only, in blue; PP2, working groups only, in purple; and PP3, household and working groups with equal weight, in green). Parameters relating to self-evaluation are set to the same values as in SE3 (see Table \ref{tab:sim_para}). Shaded areas around the solid line show standard deviation. Each simulated profile is averaged over 25 runs. The actual daily incidence between December 2021 to June 2022 is shown in scattered diamonds.}
    \label{fig:PP}
\end{figure}

We observed that adding the peer pressure weight $\lambda$ affected the number of waves and the corresponding incidence peaks (Appendix Fig. \ref{app:fig_lambda} and Section \ref{app:results}), with a mid-range $\lambda = 0.4$ providing the best match. The selection of different peer-pressure forming social contexts (HH, WG, or both) did not affect the first wave, with all scenarios demonstrating a high SD adoption in response to the prominent first peak (Fig.~\ref{fig:PP}). However, the choice of peer-pressure social contexts affected the second and third pandemic waves. Specifically, opinions inferred from a smaller social context (HH, with between 1 to 6 household members; PP1) led to a smaller variability in SD adoption after the first incidence peak, resulting in a slower decline in SD adoption and consequently, a lower second incidence peak. Opinions inferred from a larger social context (WG, up to 20 co-workers; PP2), on the other hand, accentuated fluctuations in the SD adoption, resulting in more drastic oscillatory incidence dynamics during the second and third waves. When the peer pressure originated from both contexts HH and WG (PP3), the resultant pandemic dynamics were closest to the observed incidence, showing three distinct waves: an acute, high incidence peak around day 60, followed by two smaller incidence peaks around day 130 and day 170 (see Table \ref{tab:sim_peak} for the comparison between simulated and actual incidence dynamics). 

These findings were in concordance with our previous study \cite{chang_persistence_2023}: the fluctuating adoption of social distancing strongly contributed to the recurrent waves observed during the Omicron pandemic stage in Australia. In this study, we further demonstrated that the changes in the SD adoption can themselves be attributed to several opinion-forming factors: risk aversion, memory horizon, perception fatigue, peer pressure, as well as selection of the peer-pressure forming social context(s).

\begin{table}[!t]
\renewcommand*{\arraystretch}{1.3}
    \centering
\caption{Simulation scenarios of opinion dynamics using self-evaluation and peer pressure.}
    \begin{tabular}{cccc}
        \multicolumn{4}{c}{Self-evaluation} \\
        \toprule
        simulation & $\beta$    & $\gamma$ & $T$ (days) \\
        \midrule
         SE1       & 0.5        & 0        & 7\\
         SE2       & 0.5        & 0        & following Eq. \ref{eq:sigmoid} \\
         SE3       & 0.5        & 0.002    & following Eq. \ref{eq:sigmoid} \\
         \bottomrule\\
         
        \multicolumn{4}{c}{Peer pressure} \\
        \toprule
        simulation & $\lambda$    & $\psi^\textrm{HH}$ (Household) & $\psi^\textrm{WG}$ (Working groups) \\
        \midrule
         PP1       & 0.4        & 1        & 0\\
         PP2       & 0.4        & 0        & 1 \\
         PP3       & 0.4        & 0.5      & 0.5\\
         \bottomrule
    \end{tabular}

    \label{tab:sim_para}
\end{table}

\begin{table}[!t]
\renewcommand*{\arraystretch}{1.4}
    \centering
\caption{Mean incidence peaks and the corresponding simulation day in the considered simulated scenarios.}
    \begin{tabular}{cp{2cm}p{2cm}p{2cm}}

        \toprule
        Scenario & 1st peak \newline (simulation day)    & 2nd peak \newline (simulation day) & 3rd peak \newline (simulation day) \\
        \midrule
        Actual     & 150,700 (58)       & 66,780 (128)       & 58,080 (177) \\
         SE1       &   105,500 (58)     & 71,020 (114)       & 41,680 (151)\\
         SE2       &  99,740 (58)       &  64,400 (145)       & Not formed \\
         SE3       &  120,800 (59)       & 87,420 (135)    &  Not formed\\
         PP1       &  107,900 (59)       & 42,750 (138)        & 69,690 (183) \\
         PP2       & 114,900 (59)        & 81,890 (134)       & 69,690 (183) \\
         PP3       & 122,900 (58)        & 63,210 (131)     & 45,590 (174) \\
         \bottomrule
    \end{tabular}
    \label{tab:sim_peak}
\end{table}

\section{Discussion and conclusion}
\label{sec:discussion}

Recurrent pandemic waves are a systemic phenomenon observed during spread of many infectious diseases, including during the COVID-19 pandemic \cite{campi2022sars}. One of the factors contributing to this phenomenon is the fluctuating compliance with or adoption of non-pharmaceutical interventions, such as social distancing (SD)~\cite{chang_persistence_2023}. However, the quantitative understanding of the nonlinear dependencies and feedback between the pandemic spread and the opinion-driven compliance/adoption of social distancing behaviour remains elusive. 

To address these questions we modelled the dynamics of opinions which shape the decisions on SD adoption, using a large-scale pandemic ABM. Our integrated model captured complex feedback between opinion-driven SD adoption and pandemic spread, producing a fluctuating  SD-adoption profile that generated recurrent pandemic waves. The identified SD-adoption profile is in a qualitative agreement with the actual mobility reduction during this period, observed across the workplace, retail, and transport \cite{mobility-sep-2022}. Our prior study \cite{chang_persistence_2023}  retrospectively selected six SD-adoption fractions in a sequence of five-step changes, while minimising the difference between simulated and actual incidence curves. In contrast, the opinion dynamic model proposed and validated in this study produced a more intuitive and constructive explanation, using five underlying factors, attributing the varying SD-adoption to a combined effect of individual risk aversion and social peer pressure. We identified an opinion formation scenario that best matches the recurrent waves observed during the Omicron stage in Australia by fine-tuning these five factors:  risk aversion $\beta$, fatigue $\gamma$, memory horizon $T$, peer pressure weight $\lambda$, and selected opinion-affecting social contexts $\psi_g$.

In contrast with many pandemic opinion dynamics models, our model includes a heterogeneous population, with agents interacting both in terms of both infection transmission and opinion dynamics. This study extended our pandemic ABM \cite{chang_modelling_2020,chang_persistence_2023,chang_simulating_2022,zachreson_how_2021,BMC} by incorporating re-infection and waning immunity. This addition allowed for (i) the recovered agents to return to a susceptible state 60 days after their recovery, and (ii) the immunity from vaccinations and past infections to wane over time. 
We acknowledge several limitations of this work. Firstly, we assumed that all agents use the normalised nationwide incidence (i.e., ``global'' incidence) to form the perceived risk of infection with identical risk aversion parameter $\beta$ and fatigue rate $\gamma$. This simplification did not consider scenarios where (i) agents may have different risk aversion levels and perception fatigue rates (i.e.,  $\beta$ and $\gamma$ distributed across the population), and (ii) agents may use ``local'' incidence (e.g., from the local health district, city, or state) during the risk evaluation process. Secondly, we interpreted peer pressure as the average opinion across specific social contexts where all opinions are weighted equally. In reality, however, some members of the community may have greater influence than others (e.g., opinion leaders and community champions). One possible future direction is to model several agents assigned a higher weight of opinion affecting a large social context (e.g., community) and explore the impact of these agents on the coupled pandemic and opinion dynamics. We also acknowledge that there are other methods to model peer pressure, for example, conformity~\cite{JAVARONE201419}.

We quantitatively modelled a nonlinear interplay between opinion dynamics and pandemic spread in a heterogeneous population, revealing a balanced impact of individual risk perception and social peer pressure on the SD-adoption decisions and resultant recurrent pandemic waves. We believe that a better understanding of intricate underlying factors shaping opinion dynamics during pandemics may inform policy-makers in designing more effective crisis response approaches.  
\section*{Acknowledgement}
This work was supported by the Australian Research Council grant DP220101688 (MP, QDN, SLC) and the University of Sydney's Digital Science Initiative (SLC, QDN, CS, RR, VS, TS, AM, MP). In addition, SLC is supported by the University of Sydney Infectious Diseases Institute Seed Grant 220182. AM is supported by NHMRC Investigator Grant. The simulations involved in this work used the high-performance computing cluster (Artemis) provided by the Sydney Informatics Hub at the University of Sydney.
\appendix
\section*{AMTraC-19 modelling framework}
\subsection*{Population generation}
We generated an artificial population representing the main demographic characteristics and commuting network of the Australian population, by using several high-resolution datasets, including the Australian Census of 2021~\cite{abs_general_webpage}, international air traffic reports with incoming passenger flows in Australian airports~\cite{BITRE_airport_data}, and educational registration records including the number of schools and students~\cite{ACARA_Data}. The population generation process assigned each agent with multiple demographic characteristics (e.g., age, gender, residency location) and social contexts in different settings: residential (e.g., household, household cluster, etc.); educational (e.g., classroom/school for agents aged $18$ years or younger); and workplace (for agents aged over $18$ years). While the residential contexts are determined by residential demographics, the workplace and educational contexts follow the commuting network (travel to work and travel to school respectively, depending on agents' age groups)~\cite{fair_creating_2019}. The detailed up-to-date description of the population generation process can be found in Supplementary Materials of a recent study~\cite{BMC} and the user guide of our open-source software~\cite{chang_amtract_user_guide_2022_7325675}.

\subsection*{Disease transmission}
Using the generated surrogate population, we stochastically simulate infection transmission based on interactions within the agents' social contexts. If an agent in Susceptible state is exposed to the infection within one or more of its social contexts, it may get infected (dependent on infection probabilities), and then progress through several states: Latent, Infectious (asymptomatic or symptomatic), and Recovered. We assumed no population dynamics during the simulated period (i.e., no births and deaths).   

We use the daily probability of interaction $q_{j \rightarrow i}(g)$, potentially transmitting infection from agent $j$ to agent $i$, which depends on the ages of agents $j$ and $i$, and on the context $g$ in which the interaction occurs. These probabilities $q_{j \rightarrow i}(g)$ have been defined and calibrated in previous studies \cite{chang_modelling_2020,chang_persistence_2023}.

At cycle $n$, given the interaction probabilities $q_{j \rightarrow i}(g)$, we define the transmission probability $p_{j \rightarrow i}(n,g)$ which depends on the epidemiological characteristics and natural history of the disease:
\begin{eqnarray}
\label{eq:individual_transmission_prob}
    p_{j \rightarrow i}(n,g) = \kappa \ f_j(n-n_j) \ q_{j \rightarrow i}(g)    
\end{eqnarray}
where $\kappa$ is a global transmission scalar used to calibrate the reproductive number $R_0$, $n_j$ denotes the infection onset time for agent $j$, and the agent-specific function $f_j(n-n_j)$ is the natural history of the disease, reflecting the infectivity of agent $j$ as its infection progresses. At the time cycle $n$, if agent $j$ is not infected, $f_j(n-n_j)=0$. If agent $j$ is infected, $f_j(n-n_j) \geq 0$. 

When a susceptible unvaccinated agent $i$, not protected by any non-pharmaceutical interventions, is exposed to the infection for the first time, the infection probability (i.e., probability of changing the agent state  to Latent) across all social contexts is calculated as:
\begin{equation}
\begin{aligned}
    p_i(n) &= 1 - \prod_{g \in G_i(n)} \; \prod_{j \in A_g\backslash\{i\}} \left( 1 - p_{j \rightarrow i}(n,g) \right)
\end{aligned}
\label{eq:general_infection_prob}
\end{equation}
where $G_i(n)$ denotes the set of all social contexts $g$ that agent $i$ interacts with during the time cycle $n$, $A_g\backslash\{i\}$ denotes the set of agents in $g$ (excluding agent $i$).

An infected agent can be either symptomatic or asymptomatic. The probability of symptomatic illness $z_i(n)$ is determined with respect to the infection probability $p_i(n)$, by incorporating an age-dependent scaling factor $\sigma_i$ that represents the fraction of symptomatic cases among the total cases:
\begin{eqnarray}
    z_i(n) = \sigma_i \ p_i(n)
    \label{eq:infection_prob_symptomatic}
\end{eqnarray}
where for adults ($\text{age} > 18$) $\sigma_i = \sigma^a$;  and for children ($\text{age} \leq 18$) $\sigma_i = \sigma^c$.

We calibrated the associated parameters (summarised in Table~\ref{tab_supp:epi_para}) to match the characteristics of the Omicron variant following~\cite{BMC,chang_persistence_2023}.
In this study, we also introduced re-infections, so that agents may become susceptible again upon recovery. The transition between Recovered and Susceptible states takes place after a fixed post-recovery period (e.g., 60 days), which aligns with clinical observation of re-infection periods for Omicron infections~\cite{CDC_inref, NSW_reinf}. 


\renewcommand{\arraystretch}{1.2}
\begin{table*}
    \centering
         \caption{Model parameterisation for Omicron variant adopted by AMTraC-19. The last two rows show the corresponding basic reproductive number ($R_0$) and generation/serial interval ($T_{gen}$), calibrated previously \cite{chang_persistence_2023,BMC}.}
    \begin{tabular}{lll}
    \toprule
     Model parameters         & Value & Note \\
         \midrule
         $\kappa$                        & 23  & Global transmission scalar \\
         $T^{\text{inc}}$, mean&               3 ($\mu$=1.013 $\sigma$=0.413) & Incubation period (log-normal) \\
         $T^{\text{rec}}$, mean and range      & 9 [7,11] &Recovery period, mean and range (uniform)\\
         $T^{\text{lat}}$, fixed                  & 0 & Latent period \\
         $\sigma^\textrm{a}$                     & 0.67  & Probability of symptoms (adults, age $>$ 18) \\ 
         $\sigma^\textrm{c}$                     & 0.268 & Probability of symptoms (children, age $\leq$ 18)  \\ 
         $\pi$                   & 0.12   & Daily case detection probability (symptomatic)  \\ 
         $\pi'$                 & 0.01 & Daily case detection probability (asymptomatic)  \\ 
         $R_0$, mean and 95\% CI        &  19.56 [19.12, 19.65]  &Basic reproductive number\\
         $T^{\text{gen}}$, mean and  95\% CI&  5.42 [5.38, 5.44]  & Generation/serial interval\\
         \bottomrule
    \end{tabular}
    \label{tab_supp:epi_para}
\end{table*}
\subsection*{Non-pharmaceutical interventions}
Non-pharmaceutical interventions (NPIs) may reduce the infection probability for susceptible agents. We characterise each NPI by (i) macro-distancing level, defined as the NPI-adopting fraction of all population, and (ii) micro-distancing level, defined as the adjusted interaction strengths between the NPI-adopting agents and other agents in the same social group (see Table~\ref{tab:NPI} for NPI parameterisation). 

At each time point, relevant NPIs incorporated in our model (CI, HQ, SD, and SC) affect the agents according to their state, over a certain duration $D_i$. While CI and HQ are activated at the start of the simulation and last throughout the simulated period, SD and SC are triggered only if the national cumulative incidence exceeds a certain threshold. Assignment of NPI-adopting agents follows a Bernoulli process and an agent may adopt multiple NPIs. However, the interaction strengths are adjusted to the value of $F_j(g)$ following a descending priority assignment order: CI, HQ, SD, and SC. In this study, we assumed a constant level of adoption or compliance with for CI, HQ, and SC, while the SD adoption level is formed by the opinion dynamics detailed in Section \ref{method:model}. Therefore, there are no macro-compliance or micro-duration parameters for SD.

Formally, the probability of infection for NPI-adopting agents is extended from Eq. \ref{eq:general_infection_prob} as:
\begin{equation}
\begin{aligned}
   p_i(n) =  1 - \prod_{g \in G_i(n)} \Biggl[&1 - F_i(g) \Biggl( 1 -  \\
   &  \prod_{j \in A_g\backslash\{i\}} \Bigl(1 - F_j(g) \ p_{j \rightarrow i}(n,g) \Bigr) \Biggr) \Biggr]
    \label{eq:infection_prob_compliant_agent}
    \end{aligned}
\end{equation}
where $F_j(g)$ is the strength of the interactions between agent $j$ and other agents in the social context $g$. For NPI-adopting agents $j$, the interaction strength is $F_j(g) \neq 1$. For non-adopting agents $j$, the interaction strength remains unchanged, $F_j(g) = 1$. Note that Eq.\ref{eq:infection_prob_compliant_agent} reduces to Eq. \ref{eq:general_infection_prob} if $F_j(g) = 1$.

\renewcommand{\arraystretch}{1.2}
\begin{table}[ht]
\caption{The macro-distancing parameters (population fractions) and micro-distancing (interaction strengths) for the considered NPIs. The micro-duration of CI is limited by the disease progression in the affected agent $i$,  $D_i$. }
\begin{adjustbox}{width=\columnwidth,center}
    \begin{tabular}{cccccccc}
    \toprule
         & \multicolumn{3}{c}{Macro-distancing (population fractions)} & \multicolumn{4}{c}{ Micro-distancing (interaction strengths)}\\
     Intervention &  Compliance level &  Duration (days) &  Threshold (cases)&     Household    &  Community &  Workplace/School &  Duration (days) \\
    \midrule
    CI      & 0.7 & 196 & 0 & 1.0 & 0.25 & 0.25 & $D_i$ \\ 
    HQ      & 0.5 & 196 & 0 & 2.0 & 0.25 & 0.25 & 7 \\
    $SC^{s/t}$  & 1.0 & 110 & 100 & 1.0 &0.5 &0 & 110 \\
    $SC^p$  & 0.25& 110 & 100 & 1.0 &0.5 &0 & 110 \\
    $SD$    & dynamic & 196 & 400 & 1.0&0.1 & 0.25 & dynamic \\ 
    \bottomrule
    \end{tabular}
   \label{tab:NPI}
\end{adjustbox}
\end{table}
\subsection*{Vaccination}
Vaccination reduces the infection risk for susceptible agents. In this study, we modelled a high pre-emptive vaccination coverage (90\% of the population, or 22.89 million agents), attained before the Omicron stage. The vaccine distribution follows an age-dependent proportion \cite{zachreson_how_2021}: 8.33\% of vaccinated agents are aged 18 years or younger (age $\leq 18$), 83.34\% are aged between 18 and 65 years ($18 <$ age $<65$), and 8.33\% are aged 65 years or older (age $\geq 65$). Two types of vaccines are rolled out in our model, each covering 11.443M agents (or 45\% of the simulated population): the ``priority'' vaccine with a higher clinical efficacy ($\eta^\textrm{pri}$),  and the ``general'' vaccine with a medium efficacy ($\eta^\textrm{gen}$). Table \ref{tab:re-infection_parameters} shows the efficacy values aligned with clinical observations following the booster vaccination against the Omicron variant~\cite{Andrews2022,beenstock2023joint}. 

Following \cite{zachreson_how_2021, chang_simulating_2022, chang_persistence_2023}, we decomposed $\eta$ into the susceptibility-reducing efficacy $\theta$ and the disease-preventing efficacy $\zeta$ as: $\eta = \theta + \zeta - \theta  \ \zeta$.
\subsection*{Waning immunity}
The agents with vaccination and/or past infections can develop immunity against the disease, reducing their risk of infection \cite{szanyi2022log}. 
In our model, for an agent $i$ with a history of records $H_i$ containing all vaccinations and past infections $r$ ($r \in H_i$), we assumed that the immunity from either vaccination or infection can be characterised by two separate immunity components: (i) compound immunity against symptomatic infection ($M^\textrm{c}_i$, if agent $i$ is susceptible to the virus), and (ii) immunity against forward infection ($M^\textrm{t}_i$, if agent $i$ is infected and may transmit the virus to other susceptible agents).

At simulation cycle $n$, for every past record $r$, each immunity component wanes over time as follows:
\begin{equation}
\begin{split}
    M_i^\textrm{c}(n,r) = M^\textrm{c}(r) &  [1 - \text{min}(1,(n-n_r)\epsilon )] \\
    M_i^\textrm{t}(n,r) = M^\textrm{t}(r) &  [1 - \text{min}(1,(n-n_r)\epsilon )]
    \label{eq:waning}
\end{split}
\end{equation} 
where $\epsilon$ is the immunity waning rate per simulation cycle, $n_r$ is the time cycle of past vaccination or infection record $r$, and
$M^\textrm{c}(r)$ is the peak compound immunity developed from either vaccination ($M^\textrm{c}= \eta^\textrm{pri}$ or $M^c=\eta^\textrm{gen}$), or past infection ($M^\textrm{c}=0.7$)~\cite{shrestha_coronavirus_2022, franchi_natural_2023}, while  $M^\textrm{t}(r)$ is the peak immunity against forward transmission (See Table \ref{tab:re-infection_parameters}).

For agents with multiple vaccination and/or infection records, the compound immunity against symptomatic infection, denoted $M_{i}^\textrm{c}(n,H_i)$, and the immunity against infection transmission, denoted $M_{i}^\textrm{t}(n,H_i)$, are accumulated non-linearly with an upper bound of perfect immunity of $1$:
\begin{equation}
\begin{split}
M_{i}^\textrm{c}(n,H_i) = \text{min}\left(\sqrt{\sum_{r \in H_i} \left [M_{i}^\textrm{c}(n,r) \right ]^2}, 1\right) \\
M_{i}^\textrm{t}(n,H_i) = \text{min}\left(\sqrt{\sum_{r \in H_i} \left [M_{i}^\textrm{t}(n,r) \right ]^2}, 1\right)
\label{eq:accumulation}
\end{split}
\end{equation}

Similar to the vaccine efficacy, we decomposed compound immunity $M^{\textrm{c}}_i(n,H_i)$ into susceptibility-reducing immunity ($M^\theta_i$) and disease-preventing immunity ($M^\zeta_i$).
\begin{equation}
    \begin{aligned}
        M^\zeta_i(n,H_i) = M^\theta_i(n,H_i) = 1 - \sqrt{1-M^c_i(n,H_i)}
    \end{aligned}
    \label{eq:I_s_d}
\end{equation}
The infection probability for a susceptible agent $i$, given their compliance with NPIs, the immunity levels, and the exposure to infection in all contexts $g \in G_i$, is then defined as follows:
\begin{equation}
\begin{aligned}
   & p_i(n)  = \left(1-M^\theta_i(n,H_i) \right) \left [1 - \prod_{g \in G_i(n)} \Biggl[ 1 - F_i(g)  \right. \\ 
    & \left.  \Biggl (1 - \prod_{j \in A_{g}\setminus \{i\}} \Bigl( 1 - \bigl(1 - M_j^\textrm{t}(n,H_j) \bigr)  F_j(g) \ p_{j \rightarrow i} (n,g) \Bigr) \Biggr)  \Biggr] \right] \\
    \label{eq:p_i}
\end{aligned}
\end{equation}
The agent probability to become ill (symptomatic) is influenced by the disease-preventing immunity component, ($M^\zeta_i$): 
\begin{equation}
    z_i(n) = \left(1 - M^\zeta_i(n, H_i) \right) \ \sigma_i \ p_i(n)
    \label{eq:ze_i(n)}
 \end{equation}
where $\sigma_i$ is the fraction of symptomatic cases among total infection cases (see Section \ref{app:detection} for more details). Note that Eq.~\ref{eq:p_i} reduces to Eq. \ref{eq:infection_prob_compliant_agent}, and Eq. \ref{eq:ze_i(n)} reduces to Eq. \ref{eq:infection_prob_symptomatic}, if agent $i$ has no prior immunity. Table \ref{tab:re-infection_parameters} summarises the parameters used in modelling waning immunity.
\begin{table}[!t]
    \centering
    \caption{Simulation parameters for reinfection and vaccine waning.}
   \begin{tabular}{lll}
       \toprule
        Parameter    & Value & Reference/Note \\
        \midrule
        $\eta^{pri}$ & 0.7 & Clinical efficacy, priority vaccine \\
        $\eta^{gen}$ & 0.5 & Clinical efficacy, general vaccine \\
        $M^c$       & 0.7 or 0.5& Peak immunity against symptomatic infection \\
        $M^t$       & 0.4 & Peak immunity against  infection transmission\\
        $\epsilon$     & 0.00034 &  Immunity waning rate per cycle \\
        $T^{\text{post}}$  & 60      & Post-recovery period before reinfection (days) \\
        \bottomrule
    \end{tabular}
    \label{tab:re-infection_parameters}
\end{table}
\subsection*{Disease detection}
\label{app:detection}
We assumed that only a fraction of total infections are detected on a daily basis, given the voluntary self-reporting system adopted in Australia during 2021--2022. We also assumed that the asymptomatic cases are more difficult to detect than symptomatic cases so that detection probabilities $\pi$ (symptomatic) and $\pi'$ (asymptomatic) follow $\pi \gg \pi'$.

During the infection period, the probability that an infected agent $i$ is detected within $d$ days since the infection onset is defined as follows:
\begin{equation}
    \delta_i(d) = 1 - (1 - \pi_i)^{d}
    \label{eq:p_d_i}
\end{equation}
where $\pi_i$ is the probability of detection for agent $i$ depending on their symptomaticity:  
\begin{equation}
    \pi_i = 
    \begin{cases}
        \pi & \text{if agent $i$ is symptomatic} \\
        \pi' & \text{if agent $i$ is asymptomatic} \\
    \end{cases}
\end{equation}
The expected detection probability of an infected agent within $d$ days since infection onset, accounting for all possible infections (asymptomatic or symptomatic), is then given by:
\begin{equation}
\begin{aligned}
    \mathbb{E}[\delta_i(d)] & = \sigma_i \left(1 - (1 - \pi)^{d}\right) + (1 - \sigma_i) \left(1 - (1 - \pi')^{d}\right) \\
    & = 1 - \sigma_i (1 - \pi)^{d} - (1 - \sigma_i) (1 - \pi')^{d}
\end{aligned}
    \label{eq:p_d}
\end{equation}
where $\sigma_i$ is the probability that an infection in agent $i$ is symptomatic (see Eq. \ref{eq:infection_prob_symptomatic}). We assumed that an infection is either symptomatic or asymptomatic from the onset. 

We calibrated $\pi$ and $\pi'$ as follows. According to \cite{lifeblood2022seroprevalence}, the anti-nucleocapsid antibodies prevalence elicited by natural infection was 46\% (i.e., 2,300) out of 5,000 blood donation samples collected between 9 and 18 June 2022. Arguably, this approach underestimates the cases due to the systematic difference between samples from blood donors and those from the general population. To reduce the bias, we assume that (i) blood donation samples were only collected from susceptible but not infected individuals (including those recovered from previous infection) or asymptomatic individuals and, (ii) a constant ratio between symptomatic and asymptomatic cases, $\rho= \sigma^\textrm{a}/(1-\sigma^\textrm{a})$, we use the anti-nucleocapsid antibodies prevalence to estimate the recovered population nationwide:
\begin{equation}
    \frac{2300 + \rho \times 2300  }{5000 +  \rho \times 2300 } \times 25.4M = 18.256M
\end{equation}
where the term $\rho \times 2300$ quantifies the corresponding number of symptomatic individuals present in the population at the time and not contributing to the blood donation.

Given the reported cumulative incidence of 7.3M during this period \cite{covid19data}, the fraction of detected infection cases nationwide can be approximated as: 7.3M/18.256M $\approx$ 40\%. We found that a combination of $\pi=0.12$ and $\pi'=0.01$ produced a satisfactory match at 37.8\% after six days of detection, following Eq. \ref{eq:p_d} (assuming all testing occurs within six days of infection).
\subsection*{Opinion dynamics modelling: additional results}
\label{app:results}
\begin{figure}[!t]
\vspace{1em}
    \centering
    \includegraphics[width=0.5\linewidth,trim={5cm 8cm 5cm 8cm}]{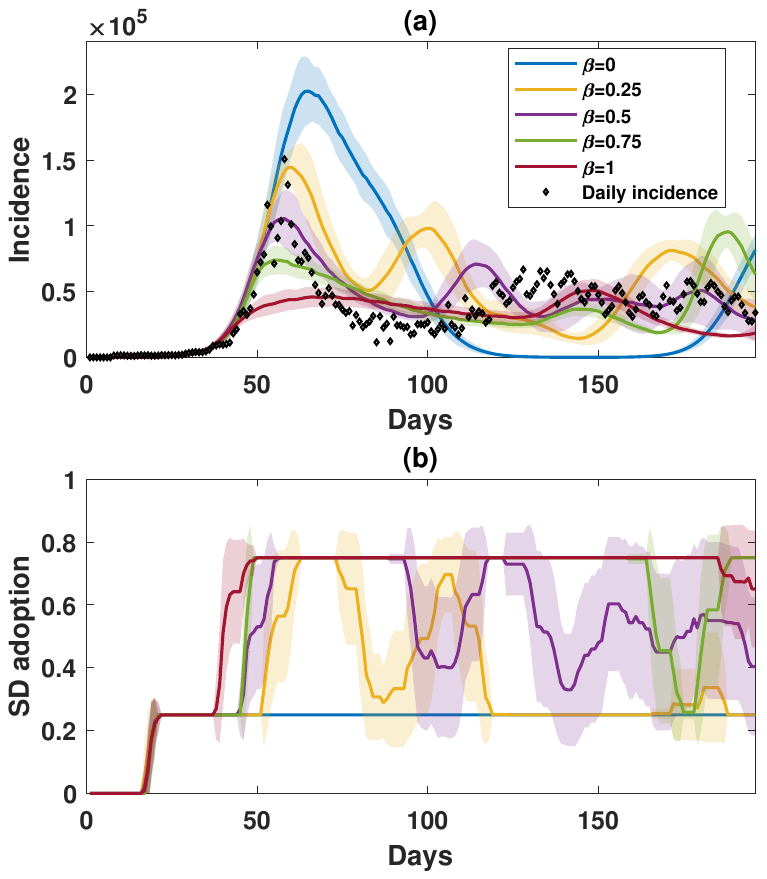}
    \vspace{1em}
    \caption{A comparison of simulated incidence (a) and SD-adoption fractions (b), with opinions produced by self-evaluation only varying $\beta=[0,1]$ with 0.25 increment. Shaded areas around the solid line show standard deviation. Each simulated profile is averaged over 25 runs. The actual daily incidence is shown in scattered diamonds.}
    \label{app:fig_beta}
\end{figure}

\begin{figure}[h]
\vspace{3em}
    \centering
    \includegraphics[width=0.5\linewidth,trim={5cm 8cm 5cm 8cm}]{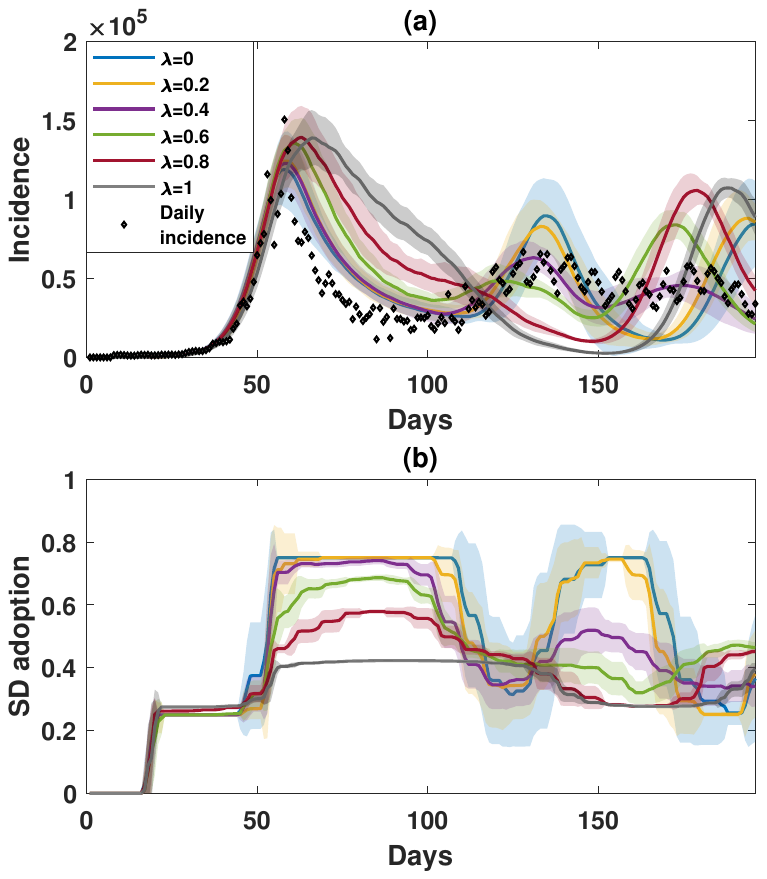}
    \vspace{1em}
    \caption{A comparison of simulated incidence (a) and SD-adoption fractions (b), with opinions produced by combining self-evaluation and peer pressure, varying the peer pressure weight $\lambda=[0,1]$ with increment 0.2. The considered social contexts are HH and WG, with equal weights ($\psi^\textrm{HH}=\psi^\textrm{WG}=0.5$). Note that $\lambda=0$ is equivalent to SE3 in Fig. \ref{fig:SE}. Shaded areas around the solid line show standard deviation. Each simulated profile is averaged over 25 runs. The actual daily incidence is shown in scattered diamonds.}
    \label{app:fig_lambda}
\end{figure}

Figure \ref{app:fig_beta} shows that risk aversion, quantified by the perceived probability of infection $\beta$, strongly influenced incidence waves and the associated peaks. While lower $\beta$ (lower risk aversion) led to more frequent changes in both SD adoption and the incidence waves, higher $\beta$ (higher risk aversion) resulted in a more stable SD adoption associated with a reduced number of waves at lower incidence peaks. We note that the extreme cases of $\beta$ failed to produce multi-wave patterns. Specifically, $\beta=0$ produced a single-wave pandemic, with the majority of the population, $75\%$, not adopting SD (Fig. \ref{app:fig_beta}, blue profile), while $\beta=1$ produced endemic-like dynamics without acute peaks, with all rational agents choosing to adopt SD, i.e., 75\% of SD-adoption (Fig. \ref{app:fig_beta}, dark red profile). 
Figure \ref{app:fig_lambda} shows that higher peer pressure $\lambda$ encouraged the opinion convergence away from SD adoption, resulting in a prolonged first wave while delaying the second wave. When $\lambda=0$ (Fig. \ref{app:fig_lambda}, blue profile), peer pressure had no impact on the perception of the infection risk and the opinions solely depended on self-evaluation. Conversely, when $\lambda=1$ (Fig. \ref{app:fig_lambda}, grey profile), the opinion formation was purely dependent on the average opinion of the selected opinion-affecting social contexts. 

\subsection*{Google mobility and online search trends data}
\begin{figure}[h]
\vspace{2em}
    \centering
    \includegraphics[width=0.75\linewidth,trim={0cm 1cmcm 0cm 2cm}]{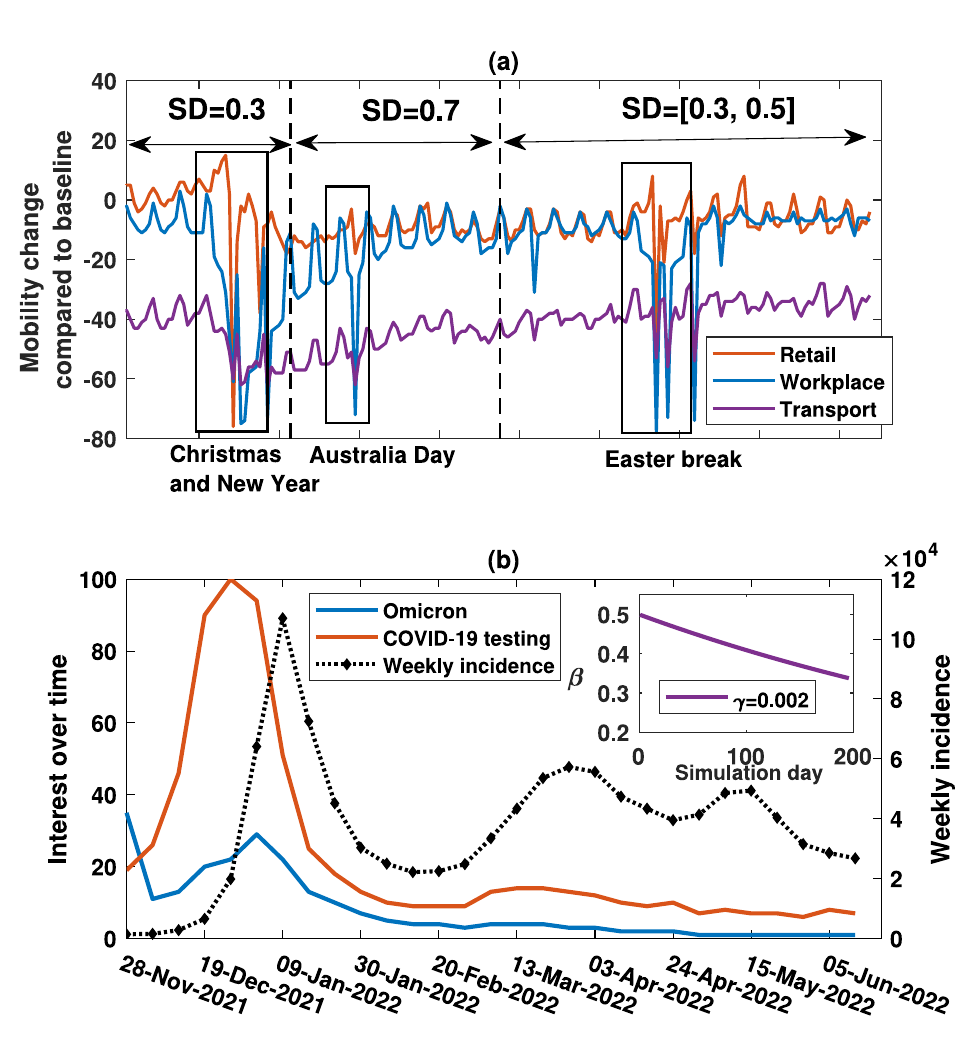}
    \vspace{1em}
    \caption{Movement trends and Google Trends during Omicron stage in Australia. (a) Movement trends compared to the baseline (the median value from the 5-week pre-pandemic period between 3 January to 6 February 2020). The vertical black lines mark the SD profile produced by opinion dynamic modelling shown in Fig. \ref{fig:PP} (PP3). Public holidays are marked in black boxes. (b) Weekly Google Trends (y-axis, left) obtained for COVID-19 related keywords. Weekly incidence is shown as dotted black line (y-axis, right). Online interest over time represents the search interest in the given period where 100 indicates the peak popularity and lower values correspond to no popularity. The inset shows how the risk aversion parameter $\beta$ declines over time given the fatigue rate $\gamma=0.002$.}
    \label{app:fig_mobility}
\end{figure}

Our simulated opinion dynamics and SD adoption fractions are in agreement with empirical evidence provided by the mobility \cite{mobility-sep-2022} and online search trends~\cite{trends-sep-2022}, reported between 26 November 2021 and 10 June 2022. Using Google COVID-19 mobility reports for workplace, retail, and transport, we note a greater mobility reduction during the period when the model generated a higher SD-adoption, indicating a reduced level of agent interactions (Table \ref{app:tab_mobility} and Fig. \ref{app:fig_mobility} (a)). In addition, we note that public interest in COVID-19 related topics declined during this time frame, as evidenced by the reduced popularity of keywords ``Omicron'' and ``COVID-19 testing'' reported in Google Trends (Fig. \ref{app:fig_mobility} (b)). This observation matched our simulating scenario where the risk aversion of infection declined over time  (i.e., declining $\beta$), governed by fatigue rate $\gamma$ (Fig. \ref{app:fig_mobility} (b), inset).

\begin{table}[!t]
\renewcommand*{\arraystretch}{1.2}
    \centering
\caption{Mean mobility reduction. See Fig. \ref{app:fig_mobility} for the time series of mobility reduction. }
\label{app:tab_mobility}
    \begin{tabular}{lp{2.5cm}p{2.5cm}p{2.5cm}}
         \toprule
         Category & SD=0.3 & SD=0.7 & SD=[0.3, 0.5] \\
                  & 26/11/21--09/01/22  & 10/01/22--05/03/22 & 06/03/22--10/06/22               \\
         \midrule
         Workplace (\%) &  -9.6    & -15.1    & -9.9      \\
         Transport (\%) &  -40.5   & -46.5    & -36.9     \\
         Retail (\%)    &  1.2     & -9.6     & -6.4      \\
         \bottomrule
    \end{tabular}
\end{table}

\clearpage
\bibliographystyle{unsrt}  


\end{document}